\documentclass{blois}

\usepackage{amsmath}
\usepackage[numbers]{natbib}
\usepackage{url}
\usepackage{hyperref}

\newcommand{\ttbar}{\ensuremath{\mathrm{t} \bar{\mathrm{t}}}}

\newcommand{\pt}{\ensuremath{p_\mathrm{T}}}

\newcommand{\TeV}{\ensuremath{{\,\mathrm{TeV}}}}
\newcommand{\GeV}{\ensuremath{{\,\mathrm{GeV}}}}
\newcommand{\fbinv}{\ensuremath{{\,\mathrm{fb}^{-1}}}}

\newcommand{\pb}{\ensuremath{\,\mathrm{pb}}}
\newcommand{\fb}{\ensuremath{\,\mathrm{fb}}}
\newcommand{\alphaS}{\ensuremath{\alpha_\mathrm{S}}}

\newcommand{\statsyst}{\ensuremath{(\mathrm{stat}+\mathrm{syst})}}
\newcommand{\stat}{\ensuremath{(\mathrm{stat})}}
\newcommand{\syst}{\ensuremath{(\mathrm{syst})}}
\newcommand{\lumi}{\ensuremath{(\mathrm{lumi})}}
\newcommand{\beam}{\ensuremath{(\mathrm{beam})}}

\newcommand{\pttop}{\ensuremath{p_\mathrm{T}^\mathrm{top}}}
\newcommand{\mtop}{\ensuremath{m^\mathrm{top}}}
\newcommand{\mtt}{\ensuremath{m^{\ttbar}}}

\newcommand{\sigmatttheo}{\ensuremath{\sigma_{\ttbar}^\mathrm{theo}}}
\newcommand{\sigmatt}{\ensuremath{\sigma_{\ttbar}}}

\begin{document}
\vspace*{4cm}
\title{Inclusive and differential results of top quark pair production from the ATLAS and CMS experiments}

\author{David Walter on behalf of the ATLAS and CMS Collaborations\footnote{Copyright 2023 CERN for the benefit of the ATLAS and CMS Collaborations. Reproduction of this article or parts of it is allowed as specified in the CC-BY-4.0 license}}

\address{European Organization for Nuclear Research (CERN), Espl. des Particules 1,\\
1211 Meyrin, Switzerland}
\conference{Presented at the $16^\mathrm{th}$ International Workshop on Top Quark Physics,\\
Traverse City, Michigan, USA, 24--29 September, 2023}

\maketitle
\abstracts{This report summarizes recent results of inclusive and differential \ttbar\ cross section measurements from the ATLAS and CMS Collaborations at the LHC. Measurements at $\sqrt{s}=7,\ 8,\ 13$, and $13.6\TeV$ are compared to state-of-the-art theory predictions, using different PDF sets, matrix element calculations, or parton shower models. No significant disagreement of a single inclusive measurement is found, with an overall trend towards lower values. For the differential measurements, no theory model is able to describe the data across all bins.}


\section{Introduction}
The top quark is the heaviest particle of the standard model (SM), has a variety of unique features such as the strongest coupling to the Higgs boson among all known particles, 
and is thus of particular interest. 
In proton-proton collisions (pp) at the CERN LHC, it is primarily produced via the strong interaction in top quark-antiquark (\ttbar) pairs. 
At the highest center of mass energy of $\sqrt{s}=13.6\TeV$, \ttbar\ events are produced in large quantities ($\approx 100$ millions at ATLAS~\cite{ATLAS} and CMS~\cite{CMS} during 2022 and 2023).
About 90\% of the \ttbar\ pairs are produced in gluons fusion, while about 10\% are produced from quark-antiquark initial states. 
As the top quark decays in almost 100\% to a b quark and a W boson, 
\ttbar\ events are categorized by the subsequent decay of the two W bosons into all-hadronic (46\%), 
single leptonic (15\% for each lepton channel), and dileptonic with the same flavored (1\%) or different flavored (2\%) leptons. 

Predictions of the inclusive \ttbar\ cross section (\sigmatttheo) up to next-to-next-to-leading order (NNLO) in perturbative quantum chromodynamics (QCD) exist, 
including soft gluon resumation up to next-to-next-to leading logarithm (NNLL)~\cite{theo}. 
Due to the high energy in \ttbar\ events and the fact that the process can be measured with high precision, similar to the precision of the theory calculations,
measurements allow stringent QCD tests. 
Results are further used in the extraction of the strong coupling constant \alphaS~\cite{top7TeVCombination,top13TeVInterpretation}, 
and in global fits to constrain the proton PDF, particularly at large momentum fractions for the gluon and the gluon to quark ratio~\cite{top13TeVInterpretation}. 
From its coupling to the Higgs boson, the top quark mass is of particular interest also for the electroweak theory (EW) 
and can also be extracted from differential cross sections~\cite{top7TeVCombination,top13TeVInterpretation}.

\section{Inclusive \ttbar\ cross section measurements at $\sqrt{s}=7\TeV$}
During Run 1 of the LHC in the years 2011 and 2012, various analyses of \ttbar\ events have been performed at $\sqrt{s}=7$ and $8\TeV$ and ATLAS and CMS,
such as the inclusive \ttbar\ cross section measurements and a combination of results from the two experiments in the dileptonic channel 
yielding a precision of $\delta \sigmatt = 2.6\%$ published in 2022~\cite{top7TeVCombination}. 
ATLAS has recently released a new \sigmatt~measurement at $\sqrt{s}=7\TeV$ using the full data set from 2011 of $4.6\fbinv$~\cite{SVM}. 
The analysis targets the single lepton final state and selects events with exactly one muon or electron, a large amount of missing energy, and three jets from which 0, 1, or 2 are required to be identified as coming from a B hadron (b jet).
Significant backgrounds arise from events with W or Z bosons in association with jets, 
from single top quark production, or processes without a prompt lepton (nonprompt background). 
The nonprompt background is estimated from data via the matrix method, also known as fake rate method. 
The probability that a nonprompt lepton is misidentified as a prompt one is first measured in an independent data set 
and then applied to events in data in a region enriched in nonprompt background. 
The obtained contribution is used as an estimate for this type of background in the signal region. 
Other backgrounds with prompt leptons are estimated using Monte Carlo (MC) simulation. 
The signal is separated from the different backgrounds using a multivariate analysis based on support vector machines (SVNs). 
SVNs are chosen because of their firm mathematical foundation, simple geometric interpretation, and robustness against overtraining. 
If a minimum of their loss function is found it is guaranteed the global minimum. 
Three SVNs are trained using 21 variables to separate \ttbar, W/Z in association with b quark, and other backgrounds, in a three-dimensional space. 
From this distribution, events are categorized in four regions and binned in one-dimensional variables. 
A bin-by-bin Poisson-based likelihood function is minimized, and systematic uncertainties in the measurement are profiled. 
The obtained cross section of $\sigmatt = 168.5 \pm 0.7 \stat^{+6.2}_{-5.9} \syst^{+3.4}_{-3.2} \lumi \pb$ has a relative uncertainty of 4\% 
and is significantly improved with respect to the 12\% precision from the previous measurement from ATLAS at $7\TeV$ in this channel.
The result is consistent with the SM prediction of $\sigmatttheo = 177^{+10}_{-11}\pb$, 
and has a tension of two standard deviations with the ATLAS measurement in the dileptonic channel.

\section{Inclusive \ttbar\ cross section measurements at $\sqrt{s}=13\TeV$}
The biggest data sample at the LHC was collected at $13\TeV$ between 2016--2018 and has since been studied in great detail. 
A new analysis from ATLAS in the dilepton channel using the full data set of $140\fbinv$ documents the most precise measurement of \sigmatt~so far~\cite{atlas_13TeV}. 
Events containing an opposite signed (OS) $\mathrm{e}^{\pm}\mu^{\mp}$ pair and either one or two b jets are selected. 
Nonprompt background contribution is estimated from data using events with the same sign (SS) $\mathrm{e}^{\pm}\mu^{\pm}$ dilepton pair using the OS to SS ratio from simulation. 
Events from $\mathrm{Z}\to\tau\tau$ with subsequent decays of the $\tau$ leptons into an electron and a muon are modeled by MC. 
The normalization is rescaled based on a measurement of $\mathrm{Z}\to \mathrm{ee}/\mu\mu$ events produced in association with one or two b jets. 
Other backgrounds, such as single top W associated (tW) production, diboson, and \ttbar\ production in association with a boson, are estimated from simulation. 
Lepton efficiencies are corrected from scale factors measured in $\mathrm{Z}\to\mathrm{ee}/\mu\mu$ events. 
To mitigate the systematic effect of the difference in the event topology in \ttbar\ events, 
an additional lepton isolation scale factor is measured from events where one of the leptons fails the isolation criteria. 
The cross section in fiducial phase space is obtained simultaneously with the b tagging efficiency from the number of events with one or two b jets.
Extrapolated to the full phase space, this results in 
\begin{equation}
      \sigmatt = 829 \pm 1 \stat \pm 13 \syst \pm 8 \lumi \pm 2 \beam \pb
\end{equation}
which is, with a precision of 1.8\%, the most precise \sigmatt~measurement to date 
and in agreement with the SM prediction of $\sigmatttheo = 832^{+20}_{-29} (\mathrm{scale})^{+23}_{-23} (m_\mathrm{top})^{+35}_{-35} (\mathrm{PDF}+ \alphaS) \pb$. 
The unprecedented precision was achieved also because of the accurate luminosity determination by ATLAS with an uncertainty of 0.83\%~\cite{lumi}. 
Other important sources of uncertainty are related to the modeling of the top quark \pt~and the tW background, and the lepton selection.

\section{Inclusive \ttbar\ cross section measurements at $\sqrt{s}=13.6\TeV$}
The first analyses performed on the new Run 3 data at $13.6\TeV$ from 2022 were \sigmatt~measurements performed by the ATLAS and CMS experiments~\cite{cmsRun2,atlasRun2}. 
Inclusive \ttbar\ cross section measurements are in particular well suited for early data due to the high rate of \ttbar\ events and their distinctive signature. 
Given that they decay in a variety of different particles, it also presents a good opportunity for data validation. 
Going from 13 to 13.6\TeV, the \ttbar\ cross section increases by about 10\%. 
With a more precise expected precision, the measurement also represents the first meaningful test of the SM at the new precision frontier. 

The CMS analysis targets final events in the single and dileptonic final states, including the selection of jets and b jets with various multiplicities. 
This allows to obtain a sufficiently large data sample already from the first 1\fbinv~of collected data in August 2022. 
It also allows the use of information from the data as much as possible, to constrain experimental uncertainties on the less studied early data. 
Background processes with nonprompt leptons in the single lepton channel are estimated using the ABCD method, by constructing three regions orthogonal to the signal region. 
The probability of a nonprompt lepton to be misidentified as a prompt one is measured in events with exactly one jet 
and applied in regions with a similar selection than the signal region, but with requiring the lepton to fail the isolation requirement.
After a preliminary result was shown in last year's top conference, about one month after data taking, the final result has been published with a few refinements. 
For example, a new jet reconstruction algorithm based on the pileup per particle probability identification (PUPPI) is applied. 
A likelihood fit binned in the lepton flavor and multiplicity, and the multiplicity of jets and b jets yields a result of $\sigmatt = 881 \pm 23 \statsyst \pm 20 \lumi \pb$ with a precision of 3.5\%, in agreement with the SM prediction of $924^{+32}_{-40} \pb$. 
The total uncertainty also profits from a large reduction of the luminosity uncertainty 
where CMS makes use of Z boson rate measurements to validate the linearity and stability of conventional methods~\cite{zcountingDP}. 
This method has also been studied in more detail with a full uncertainty assessment based on 2017 data~\cite{zcounting} 
and is a valuable new input in precision luminosity determination in the future.

A different strategy was chosen by the ATLAS Collaboration, which targets events in the more clean dileptonic channel. 
This required a large amount of data which was initially 11.3\fbinv~and updated with the full data from 2022 of 29\fbinv. 
The final result also features the use of the jet vertex tagger to reject jets from pileup.
Similar to the 13\TeV\ analysis by ATLAS~\cite{atlas_13TeV}, events with one or two b jets are categorized and \sigmatt~is measured simultaneously with the b tagging efficiency. 
The obtained cross section of $\sigmatt = 850 \pm 3 \stat \pm 18 \syst \pm 20 \lumi \pb$ shows a small trend towards a smaller value than predicted.
In the analysis, a simultaneous measurement of $Z\to\mathrm{ee}/\mu\mu$ is performed 
and the \ttbar\ to Z cross section ratio is obtained in a consistent treatment of systematic uncertainty where the effect from luminosity largely cancels. 
The ratio is in particular interesting to test different PDF sets, where in general good agreement is observed. 

In comparison, despite having two largely different approaches, the two experiments end up with a very similar uncertainty where the luminosity has the largest impact,
followed by the uncertainty on the leptons and signal modeling. 
For CMS, also the uncertainty on b jets is significant, as expected from including the single lepton channel where the b jet requirement is needed to separate the backgrounds. 
In both analyses, likelihood fits are performed, while {ATLAS} has smaller prefit uncertainties and constraints, 
CMS starts with larger prefit uncertainties that get pulled and constrained by the data.

\section{Measurements of jet substructure variables in \ttbar\ events at $\sqrt{s}=13\TeV$}
The understanding of the jet substructure, in particular for b jets and large radius (large-R) jets from hadronic decaying top quarks, 
is crucial for analyses targeting final states with top quarks. 
Information can be used in tuning studies to improve the modeling of MC simulation which in turn can lead to a reduction of the related systematic uncertainties in measurements. 
Jet identification algorithms based on MVA techniques can be further developed by exploiting observed features.
ATLAS published a new result with differential cross section measurements of several variables related to the substructure of events with large-R jets. 
Variables related to the energy flow in jets, such as the Les Houches angularity which is a measure of the broadness 
and shows a separation between jets from quarks and gluons are studied.
Other variables are related to the 3 prong nature of large-R jets from top quarks, such as the N-jettiness. 
Single lepton (all hadronic) \ttbar\ events are selected that have at one (two) large-R jets with $\pt>350\GeV$ (and $\pt>500\GeV$). 
The matrix (ABCD) method is exploited to estimate the nonprompt (multijet) backgrounds in the single lepton (all hadronic) channel. 
For selected events, a data excess of 15 to 20\% is observed, compatible with the known mismodeling of \ttbar\ events in the regime with large top quark \pt~\cite{substructure}. 
The differential cross sections are extracted at particle level using an iterative Bayesian unfolding, 
their uncertainty is dominated by parton shower modeling, and experimental jet energy scale and resolution. 
Additionally to the one-dimensional variables, two-dimensional distributions are presented to study the modeling at different values of \mtop~and \pttop. 
A quantitative and qualitative assessment for different models is presented using $\chi^2$ tests. 
Variables found to be sensitive to the three-prong structure of the large-R jet from the top quark indicate tension between the data and some of the models. 
For instance, a reduction of the final state radiation (FSR) scale is favored.

\section{Differential \ttbar\ cross section measurements at $\sqrt{s}=13\TeV$}
Both ATLAS and CMS have recently analyzed single and multi-differential \ttbar\ cross section measurements in the dileptonic channel using the full Run 2 data of 140 and 138\fb, respectively~\cite{atlas_13TeV,cms_differential}. 
The ATLAS measurement is documented together with the previously discussed inclusive cross section measurements at 13\TeV\ and the same strategy is used. 
Instead of extracting the inclusive event numbers, the event numbers are extracted in bins of the observable to be measured 
and then unfolded to particle level using a bin-by-bin procedure. Eight variables of the lepton kinematics are obtained and compared to predictions 
using different matrix element generators and parton shower models. 
At CMS, also events with the same flavor dilepton pairs are included. 
Two different versions of event reconstruction are performed to obtain the kinematics of the two top quarks in the event, 
the so-called full and loose kinematic reconstructions. 
The full kinematic reconstruction uses hereby six constraints (two W masses, 2 top quark masses, and the conservation of \pt~in the event), 
while the top quark mass constraint is released in the loose version. This allows to measure \mtt\ in an unbiased way. 
In general, the uncertainty was reduced by a factor of around two with respect to a previous result. 
The dominating effects are from jet energy scale, backgrounds, and the lepton selection. 
The unfolding is performed to parton and particle level using a $\chi^2$ fit. 
Results are compared to NLO event generators, beyond NLO predictions, and different PDF sets. 
While the known feature of overestimation at higher \pttop~is visible for NLO predictions, 
it can be seen that beyond NLO predictions, in particular, approximate $\mathrm{N}^3$LO have a better modeling of \pttop. 
Further insight into this issue is provided from two-dimensional distributions. 
For example, it is shown that the \pttop\ mismodeling is enhanced at large values of \mtt\ and less strong for events with two or more jets. 
The analysis also presents three-dimensional distributions which are promising inputs for future PDF fits. 
In summary, both, the CMS and ATLAS differential measurements show that none of the models used for comparison can describe the data in all bins. 
While the modeling improves with higher order predictions, some bins remain to disagree.

\section{Summary}
\begin{figure}[!hbtp]
\centering
    \includegraphics[width=1.0\textwidth]{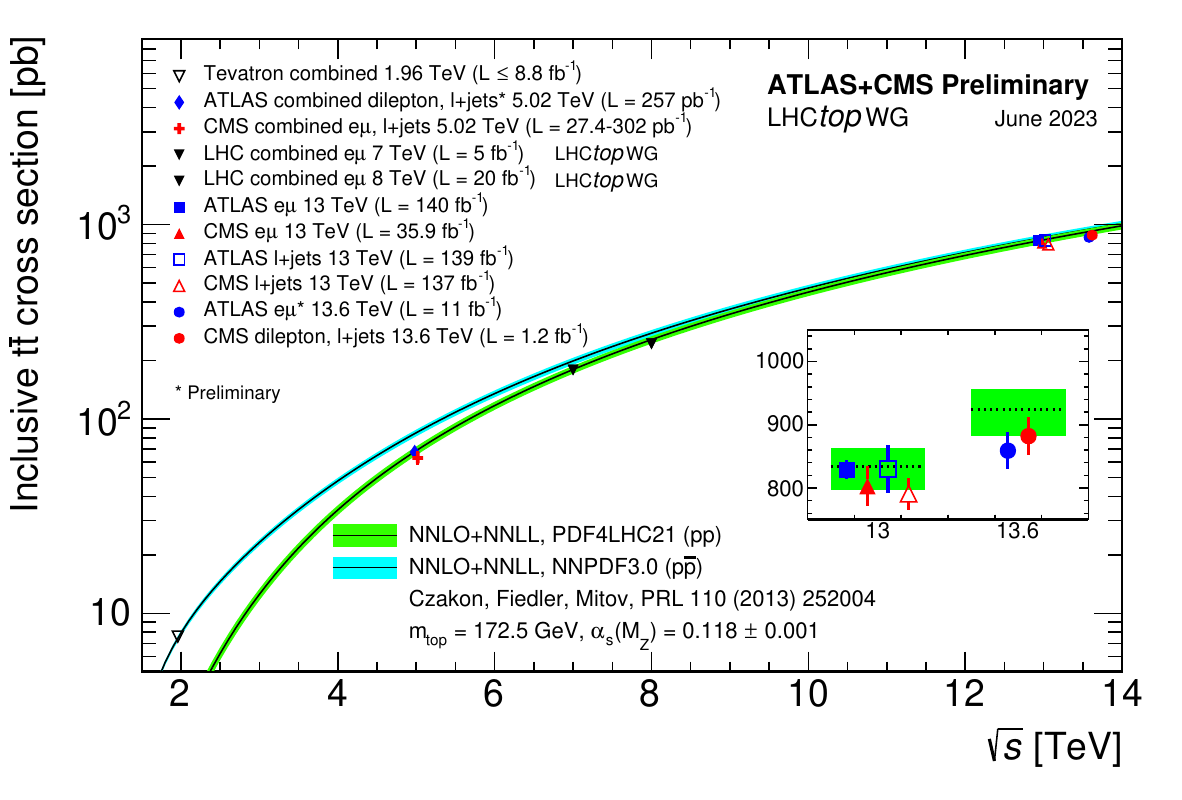}
    \caption{The \sigmatt~at varying center-of-mass energies at the LHC and the Tevatron. The experimental measurements are compared to theoretical predictions at NNLO in QCD with NNLL resummation. The theoretical predictions include uncertainties from renormalization and factorization scales, PDFs, and \alphaS. Both the measurements and theoretical calculations are reported for a top quark mass (mtop) of 172.5\GeV. To enhance clarity, measurements taken at the same center-of-mass energy have been slightly offset~\cite{plot}.}
\label{fig:summary}
\end{figure}

With the first \sigmatt~measurements at 13.6\TeV, the \ttbar\ cross section is now measured at six different centers of mass energies. A summary of inclusive \ttbar\ cross-section measurements at different centers of mass energies is presented in Fig.~\ref{fig:summary}. The \ttbar\ measurements are on the forefront of precision measurements at the LHC and the first measurement with a precision below 2\% has been achieved. While each inclusive cross section measurement has in general a good agreement with the prediction, there is a slight systematic trend towards an overprediction visible. Double and multi-differential cross section measurements have been performed and give more insight into different modeling aspects. It is observed that none of the generators can describe the data in all bins, while the disagreement increases in extreme phase space regions, and when looking in higher dimensional distributions. Higher-order predictions lead to a better agreement, but still fail to describe the data in some bins. It remains to be seen from more data, more precise measurements, and a better understanding of the theory predictions, if a real deviation from the standard model is present.

\bibliography{references}

\end{document}